%% file: main.tex
\documentclass[sn-mathphys,Numbered]{sn-jnl}% Math and Physical Sciences Reference Style
\usepackage{graphicx}%
\usepackage{multirow}%
\usepackage{amsmath,amssymb,amsfonts}%
\usepackage{amsthm}%
\usepackage{mathrsfs}%
\usepackage[title]{appendix}%
\usepackage{xcolor}%
\usepackage{textcomp}%
\usepackage{manyfoot}%
\usepackage{booktabs}%
\usepackage{algorithm}%
\usepackage{algorithmicx}%
\usepackage{algpseudocode}%
\usepackage{listings}%
\usepackage[utf8]{inputenc}
\usepackage{graphicx, nicefrac}% Include figure files
\usepackage{bm}% bold math
\usepackage{amsmath,MnSymbol,nicefrac}
\usepackage{lineno} % for initial submission
\theoremstyle{thmstyleone}%
\theoremstyle{thmstyletwo}%

\theoremstyle{thmstylethree}%


\begin{document}

\title{Design and Performance of a Magnetic Bottle Electron Spectrometer for High-Energy Photoelectron Spectroscopy}

\input{authors.tex}

\date{}

\abstract{

% \begin{linenumbers} % for initial submission
We describe the design and performance of a magnetic bottle electron spectrometer~(MBES) for high-energy electron spectroscopy.
    Our design features a ${\sim2}$~m long electron drift tube and electrostatic retardation lens, achieving sub-electronvolt (eV) electron kinetic energy resolution for high energy (several hundred eV) electrons with close to 4$\pi$ collection efficiency.
    A segmented anode electron detector enables the simultaneous collection of photoelectron spectra in high resolution and high collection efficiency modes.
    This versatile instrument is installed at the TMO endstation at the LCLS x-ray free-electron laser (XFEL).
    In this paper, we demonstrate its high resolution, collection efficiency and spatial selectivity in measurements where it is coupled to an XFEL source.
    These combined characteristics are designed to enable high-resolution time-resolved measurements using x-ray photoelectron, absorption, and Auger-Meitner spectroscopy.
    We also describe the pervasive artifact in MBES time-of-flight spectra that arises from a periodic modulation in electron detection efficiency, and present a robust analysis procedure for its removal.

% \end{linenumbers}
} 

\maketitle

\section{Introduction} \label{introduction}
    Electron spectroscopy is a powerful tool for probing the electronic and molecular structure of gaseous and condensed phase systems.
    When used in combination with ultrashort light pulses, the technique offers a route to tracking ultrafast electronic and nuclear dynamics.
    Time-resolved photoelectron spectroscopy (TRPES) of valence electrons is an established technique for probing photo-initiated molecular dynamics on the femtosecond timescale\cite{stolow_femtosecond_2004,suzuki_femtosecond_2006,neumark_time-resolved_2001,schuurman_time-resolved_2022}.
    The binding energies of core orbitals can be particularly sensitive to the local chemical environment, so the nascent ability to time-resolve small binding energy shifts in inner shells promises to be a useful tool for tracking dynamics in molecular systems \cite{wernet_communication_2017,brause_time-resolved_2018,mayer_following_2022,gabalski_time-resolved_2023,al-haddad_observation_2022}.
    Resonant x-ray absorption measured using Auger-Meitner spectroscopy also offers a powerful probe of the transient localized valence electron density in molecules \cite{mcfarland_ultrafast_2014,wolf_probing_2017,barillot_correlation-driven_2021}.
    
    A magnetic bottle electron spectrometer (MBES) combines a strong (${\sim}$1~T), inhomogenous magnetic field at the interaction point with a weaker (${\sim}$1.5~mT) uniform magnetic field through a long drift region~(or flight tube).
    This configuration achieves a high-collection efficiency for electrons while maintaining a good kinetic energy resolution.
    Electrons emitted in different directions from the interaction point are guided $via$ the Lorentz force along the magnetic field lines and into the flight tube, where they undergo betatron oscillations about the flight tube axis as they travel towards the detector.
    The original MBES design afforded 2$\pi$ solid angle collection volume, by collecting all electrons emitted in the hemisphere directed toward the flight tube~\cite{kruit_magnetic_1983, tsuboi_magnetic_1988}.
    Later developments increased the collection efficiency to 4$\pi$ by exploiting the magnetic mirror effect~\cite{cheshnovsky_magnetic_1987}.
    Some designs~\cite{radcliffe_experiment_2007,mucke_performance_2012,hikosaka_high-resolution_2014} make use of an electrostatic retardation lens to reduce the electron velocity through the flight tube and so increase kinetic energy resolution.
    The high detection efficiency of the MBES is particularly advantageous for electron-electron coincidence or covariance measurements~\cite{eland_complete_2003,frasinski_dynamics_2013,zhaunerchyk_disentangling_2015,kornilov_coulomb_2013}, for which the signal-to-noise is critically dependent on collection efficiency~\cite{frasinski_covariance_2016}.
    Finally, MBES has proven particularly useful for electron spectroscopy at x-ray free-electron lasers~(XFELs) \cite{frasinski_dynamics_2013,wolf_probing_2017,squibb_acetylacetone_2018,kornilov_coulomb_2013,pathak_tracking_2020}.

    In this work, we describe the design and performance of a MBES, which is installed at the time-resolved molecular and optical sciences~(TMO) instrument at the Linac Coherent Light Source~(LCLS)~\cite{walter_time-resolved_2022}.
    Our design features a ${\sim}2$~m long flight tube, an electrostatic retardation lens and a segmented anode electron detector, which enables the simultaneous acquisition of electron spectra in two different collection modes: high resolution and high collection efficiency.
    We show that the high collection efficiency enables covariance mapping of photoelectron/Auger-Meitner electron emission in the core ionization of nitrous oxide.
    We employ spectral domain ghost imaging~\cite{driver_attosecond_2020,li_two-dimensional_2021,wang_photon_2023} in combination with the single-shot spectral diagnostics available at the beamline to isolate the resolution of our spectrometer from the inherent spectral width associated with a self-amplified spontaneous emission~(SASE) XFEL.
    We achieve an electron energy resolution of $\frac{\delta E}{E} \sim \nicefrac{1}{100}$.

\section{Instrument Design}\label{sec:Design}

    \begin{figure*}
            \includegraphics[width =\textwidth,height =\textheight,keepaspectratio]{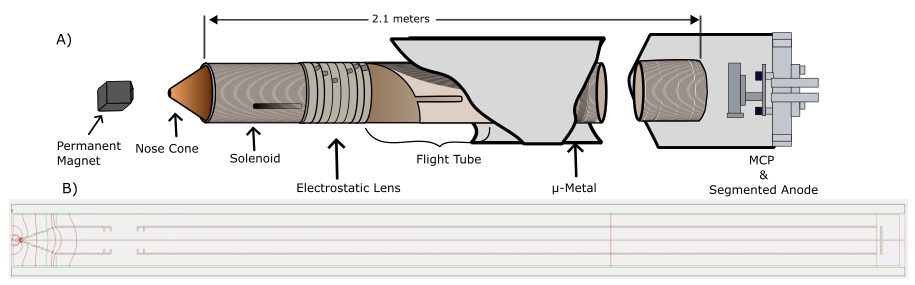}
    \caption{(A) Schematic of the MBES. Not to scale; the length of the flight tube has been truncated in this depiction. (B) To-scale cross-section of MBES taken from the graphical representation in the SIMION software package. Contour lines show regions of equivalent magnetic field strength.}
   \label{MBES_layout}
    \end{figure*}
    
    A schematic of the MBES is shown in Fig.~\ref{MBES_layout}.
    The inhomogenous magnetic field at the interaction point is generated by a nickel-coated neodynium permanent magnet.
    A soft iron cone is mounted on the permanent magnet and acts as a magnetic shunt to focus the magnetic field lines at the interaction point. 
    The permanent magnet is mounted to a remotely controlled three-axis manipulator to facilitate spectrometer alignment.
    The tip of the iron cone sits 6~mm opposite a copper nose cone, which is mounted on a $1.95$~m long drift tube.
    The interaction point sits directly between the tip of the iron cone and the tip of the copper nose cone. 
    The key elements of our design are modeled using the SIMION software (version 8.1) to simulate the trajectories of electrons through the MBES.
    These simulations are detailed throughout the results of the article. 
    
    The drift tube has an outer diameter of $76$~mm and is wrapped by kapton-insulated, $22$~AWG wire, which acts as an in-vacuum solenoid, with ${\sim}1.2$~turns/mm.
    To achieve a uniform magnetic field of $\sim1.5$~mT in the drift tube, a current of ${\sim} 1$~A is applied to the coil. 
    The drift tube is mounted inside a stainless-steel vacuum tube with  DN160~CF flanges on each end, which is surrounded by a sleeve of high-magnetic-permeability metal to reduce the effects of external magnetic fields on electrons traveling through the tube by a factor of roughly 50.
    
    There is an electrostatic lens in the drift tube to create a decelerating electric field to slow the electrons, which increases their time-of-flight, improving the energy resolution of the spectrometer.
    The position of the lens was chosen such that the magnetic field lines created by the combined fields of the permanent magnet and solenoid are parallel to the electric field lines of the lens.
    The lens begins ${\sim}183$~mm from the interaction point and consists of two pairs of three electrostatic lens plates with ${\sim}6.1$~mm  spacing, separated by ${\sim}30$~mm.
    Each set of three plates is electrically connected and a retarding effect is produced on the electrons by applying a potential between the two triplets.
    
    Electrons are detected at the end of the drift tube by a $40$~mm diameter micro-channel plate~(MCP) detector coupled to a conical anode~(Surface Concept GmbH).
    The anode is segmented into two concentric sections of 3~mm and 40~mm diameter, with a 1~mm spacing ring between them.
    In operation, a ${\sim}$1.8~kV potential difference is applied across the chevron MCP stack with a further ${\sim}300$~volts to the anode. 
    The voltages on the two anodes are decoupled from the high voltage source and amplified~(Ortec 9306, $1$~GHz preamplifier), digitized with $168$~picosecond precision (Abaco Systems FMC134 analog-to-digital card), and read into our data recording system.
    
    The segmented anode design was chosen to enable the simultaneous collection of spectra with high energy resolution (lower collection efficiency) and high collection efficiency (lower energy resolution).
    In a MBES, the time-of-flight distribution of monoenergetic electrons features a characteristic long tail, consisting of electrons emitted perpendicular to the time-of-flight axis but still directed down the flight tube by the inhomogenous magnetic field \cite{kruit_magnetic_1983}.
    In simulation, we find that the electrons with longer time-of-flight impinge on the detector at larger radii (i.e. further from the center of the detector), as shown in Fig.~\ref{fig:segmented_anode}~(A)-(C).
    The segmented anode enables the exclusion of this long time-of-flight tail by selectively detecting electrons which land close to the center of the detector.
    For measurements requiring higher detection efficiency, it is possible to also include electrons detected on the outer anode.
    Inclusion of these electrons causes a broadening of the peaks in time-of-flight, to the detriment of the kinetic energy resolution.

\section{Instrument Performance} \label{performance}

    The MBES is currently installed in the LAMP~\cite{osipov_lamp_2018} chamber at the TMO endstation~\cite{walter_time-resolved_2022} at the Linac Coherent Light Source~(LCLS) XFEL.
    In this work we characterize the performance of the MBES for x-ray photoelectron spectroscopy with an XFEL source, making use of SASE pulses with a duration of $<50$~fs and a median pulse energy of approximately $50$~$\mu$J.
    X-rays were focused to a ${\sim} 1$~$\mu$m diameter at the interaction point of the MBES using a pair of Kirkpatrick-Baez mirrors~\cite{seaberg_x-ray_2022}. 
    The base pressure of the chamber housing the MBES was below $2\times10^{-9}$ Torr and the sample gas was introduced to the interaction point using an effusive gas needle.
    The TMO endstation also features a transmissive Fresnel zone plate-based x-ray spectrometer which enables a shot-to-shot characterization of the incoming x-ray spectrum~\cite{larsen_compact_2023}.
    This measured x-ray spectrum is used for the spectral-domain ghost imaging analysis described in section~\ref{energy_resolution}.

    \subsection{Segmented Anode} \label{segmented_anode}

    \begin{figure*}
        \centering
        \includegraphics[width =\textwidth,height =\textheight,keepaspectratio]{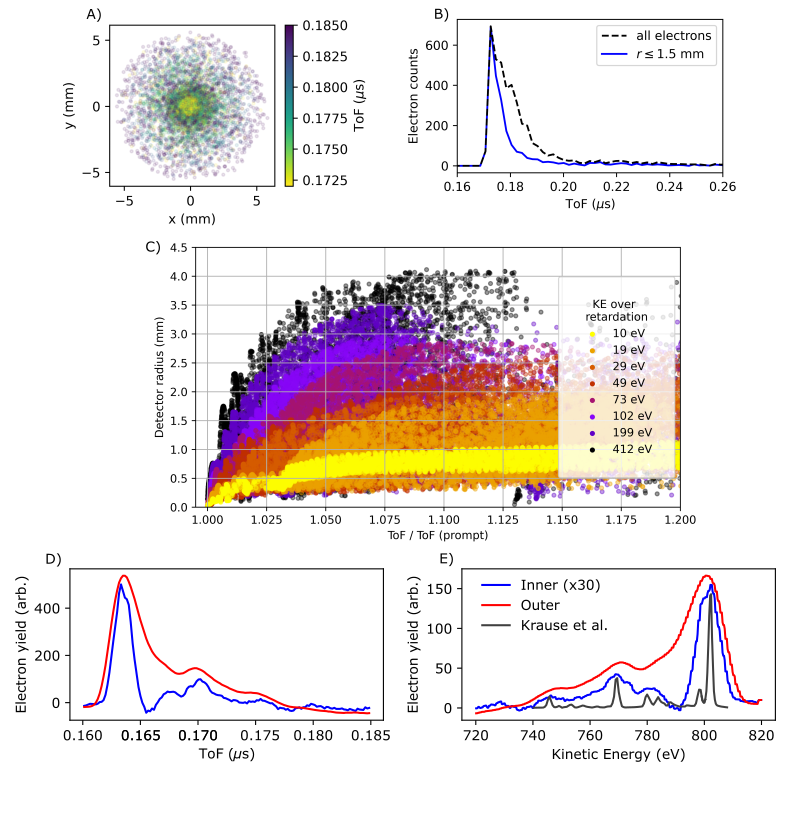}
        \caption{ 
        Motivation and performance of the segmented anode detector.
        (A) Simulated spatial distribution at the detector plane of 10,000 electrons. The electrons' time-of-flight (ToF) is encoded in the color of the point. The initial kinetic energy of the electrons is ${\sim}812$~eV and the retardation voltage is 400~V. 
        (B) Time-of-flight spectrum for the electrons shown in panel~(A). The blue curve shows the ToF spectrum discriminating on electrons landing within 1.5~mm of the center of the detector. 
        (C) Dependence of ToF on radius at which electrons impinge on the detector for multiple kinetic energies for a retardation of 100~V.
        (D) Experimental ToF spectrum for neon $KLL$ Auger-Meitner electrons produced following ionization by ${\sim}$1.35~keV x-rays, recorded with a retardation voltage of $400$~V. The red curve is for electrons detected on the outer anode and the blue curve is for electrons detected on the inner anode. The signal for the inner anode has been scaled by a factor of 30. 
        (E) Kinetic energy representation of spectra in panel (D), compared to the spectrum produced by ${\sim}$1.5~keV photons obtained using an electrostatic analyzer  ~\cite{krause_multiple_1970}.}
        \label{fig:segmented_anode}
    \end{figure*}

    As described in section~\ref{sec:Design}, the MBES features a segmented anode to discriminate the time-of-flight spectrum based on the position of electron impact on the MCP detector.
    The central part of the anode has a diameter of 3~mm, and is separated from the outer anode (diameter 40~mm) by a ring of 1~mm thickness.
    Panels (A)--(C) in Fig. ~\ref{fig:segmented_anode}~ illustrate the motivation for this design.
    Panels (A) and (B) show simulation results for 10,000 electrons sampled from an isotropic distribution of emission angles, with an initial kinetic energy of $812$~eV and $400$~V retardation applied to the electrostatic lens.
    The coupling between electron time-of-flight and arrival position at the detector plane is shown in Fig.~\ref{fig:segmented_anode}~(A). 
    The simulations show that electrons contributing to the prompt peak in the time-of-flight distribution, which are initially emitted along the axis of the flight tube, impinge close to the center of the detector.
    The slower electrons, which produce the tail in the time-of-flight distribution and are emitted with a significant velocity component perpendicular to the time-of-flight axis, impinge on the detector at larger radii.
    This effect is made clear in Fig.~\ref{fig:segmented_anode}~(B), which shows the simulated time-of-flight spectrum for all electrons (black dotted line), and only those which arrive within a 1.5~mm of the detector center (blue line).
    In panel~(C) we explore the coupling between time-of-flight and detector position for different kinetic energies and find that it is a general effect: electrons which impinge at larger radii have a longer time-of-flight.
    The retardation voltage for the simulations in panel~(C) was 100~V but we have verified that this behavior persists for all retardation voltages.

    We experimentally investigate the performance of the segmented anode detector in panels~\textbf{D} and~\textbf{E} of Fig.~\ref{fig:segmented_anode}, which show the Auger-Meitner electron spectrum of neon ionized by $1060$~eV x-rays.
    The plot shows the measured electron distributions for the same shots, measured by the inner and outer anodes.
    While the outer anode has a significantly higher electron count rate, which is attributed to its much larger surface area, the energy resolution is degraded compared to the spectrum measured on the inner anode.
    In contrast, the resolution of the Auger-Meitner spectrum measured by the inner anode is improved but has a reduced overall count rate.
    We note that the performance of the segmented anode strongly depends on the alignment of the spectrometer.
    We achieve good alignment by maximizing the electron yield on the inner anode, by translating the permanent magnet using the motorized three-axis stage on which it is mounted.

    \subsection{Energy Calibration} \label{energy_calibration}
    We calibrate the MBES by the numerical calculation of electron trajectories using SIMION.
    We simulate the trajectories for electrons emitted over an isotropic distribution of emission angles, and fit the known initial kinetic energy to the simulated time of flight for different retardation voltages.
    To account for potential deviations from the standard $KE\propto\nicefrac{1}{\left( t-t_0 \right)^2}$ behavior of a regular time-of-flight spectrometer, and an additional dependence of the calibration on the retardation voltage, we fit for a mapping from time-of-flight to kinetic energy over retardation which takes the form:
    \begin{align}
    KE(t) = \sum_{i=0}^5p_n\left( \frac{1}{\left( t-t_0 \right)^{2}} \right)^n,
    \end{align}
    where $t-t_0$ is the time-of-flight of the electron referenced to its arrival time of the x-ray pulse at the interaction point, $t_0$ and $p_n$ is a function of retardation voltage $V_R$:
    \begin{align}
    p_n = \sum_{i=0}^6 q_i^{\left( n \right)}\left( V_R \right)^i.
    \end{align}
    The dominant contribution to the fit comes from $q_1^{\left( 0 \right)}$, which indicates the behavior of the MBES is close to an ideal time-of-flight spectrometer where $KE\propto\nicefrac{1}{\left( t-t_0 \right)^2}$, and that the mapping from time-of-flight to kinetic energy over retardation is largely independent of the value of the retardation voltage.
    The energy calibration agrees extremely well with our measurements, validating the accuracy of our numerical calculations for modeling the MBES.

\subsection{Collection Efficiency} \label{sec:collection}

    \begin{figure}
        \centering \includegraphics[width=0.45\textwidth,height=\textheight,keepaspectratio]{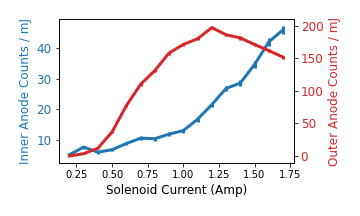}
        \caption{Yield of nitrogen $KLL$ Auger-Meitner electrons from N$_2$O measured on the inner and outer anodes, as a function of applied solenoid current. 
        }
        \label{fig:yield_coil}
    \end{figure}

    The collection efficiency of a MBES strongly depends on the kinetic energy of the electrons, the strength of the magnetic fields, and the retardation voltage applied to the electrostatic lens.
    Fig.~\ref{fig:yield_coil} shows how the magnetic field strength of the solenoid influences the electron counts measured on the inner and outer anodes.
    This measurement was performed using nitrogen $KLL$ Auger-Meitner and valence-shell photoemission from nitrous oxide~(N$_2$O) at a photon energy of $510$~eV.
    A retardation potential of $300$~V was applied to the electrostatic lens.
    The magnetic field strength was controlled by scanning the current applied to the solenoid coil. 
    As the magnetic field strength of the solenoid~(applied current) is increased, we observe an approximately linear increase in the detected electron counts on the outer anode~(red curve in Fig.~\ref{fig:yield_coil}), which reaches a maximum at ${\sim}1$~A.
    At this field strength, the cyclotron radius of the highest energy electrons emitted perpendicular to the spectrometer axis is small enough that almost all of the emitted electrons are directed onto the detector.
    As the current is further increased the electron trajectories are squeezed closer to the center of the flight tube and more electrons impinge on the inner anode, resulting in a corresponding decrease of the electron count rate on the outer anode.
    
    \begin{figure*}
        \centering
        \includegraphics[width =\textwidth,height =\textheight,keepaspectratio]{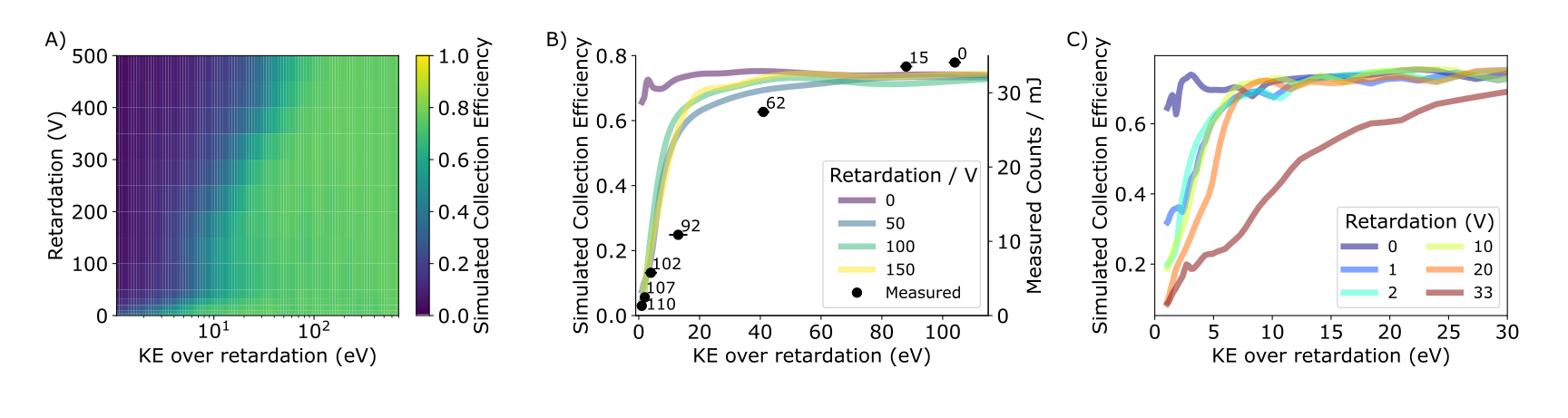}
        \caption{
        (A) Simulated values of the detection efficiency as a function of retardation voltage and kinetic energy over retardation. 
        (B) Comparison of simulated detection efficiencies from panel (A) with experimentally measured yield of ${\sim}$100~eV photoelectrons measured at different retardation voltages (black points).
        (C) Simulated electron transmission from panel (A) for lower retardation voltages and electron kinetic energies.
        }
        \label{fig:det_eff_lineout}
    \end{figure*}

    We also studied the effect of the voltage applied to the retardation lens on the MBES collection efficiency.
    SIMION simulations show that the collection efficiency is almost independent of electron kinetic energy when no retardation field is applied to the electrostatic lens. 
    Once a retarding field is applied, the collection efficiency \textit{vs.} kinetic energy over retardation (i.e. the final kinetic energy of the electron through the flight tube) has a weak dependence on the applied retardation voltage.
    This behavior is illustrated in Fig.~\ref{fig:det_eff_lineout}~(A).  
    
    We validate the performance of our SIMION model using the measured yield of ${\sim}100$~eV nitrogen $K$-shell photoelectrons ionized from N$_2$O by $510$~eV x-rays. 
    The black points in Fig.~\ref{fig:det_eff_lineout}~(B) show the electron yield as a function of kinetic energy over retardation.
    This data was collected by scanning the applied retardation voltage while keeping the x-ray photon energy fixed at $510$~eV. 
    The colored lines show the simulated detection efficiency for different retardation voltages selected from panel~(A).
    In both experiment and simulation, we observe a steep decrease in collection efficiency as the retardation voltage approaches the initial kinetic energy of the electron. 
    Panel~(C) shows the region of panel (A) corresponding to retardation voltages below $50$~V.
    At lower retardation voltages and kinetic energies, the detection efficiency depends more strongly on the retardation voltage.

    \subsection{Electron-electron covariance}

    \begin{figure}
        \centering
        \includegraphics[width =0.5\textwidth,height =\textheight,keepaspectratio]{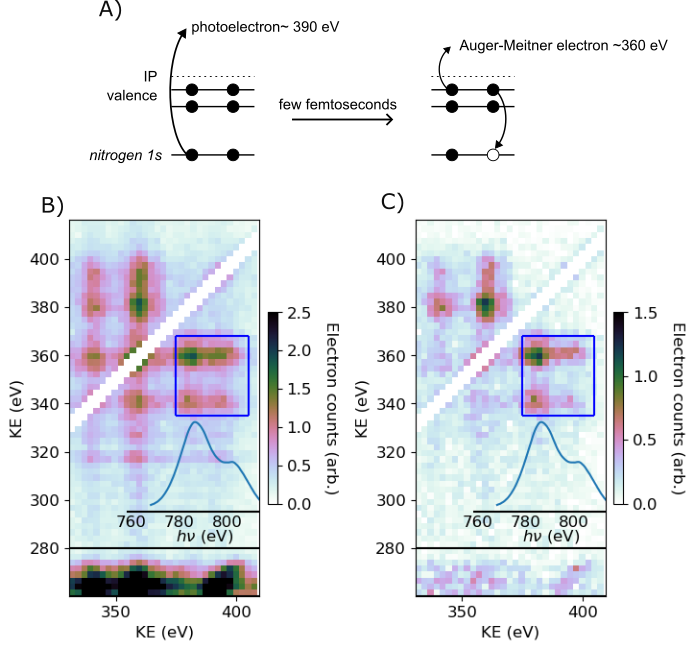}
        \caption{(A) Schematic of Auger-Meitner decay following nitrogen $K$-shell photoionization of N$_2$O by ${\sim}$790~eV x-rays. (B) Region of covariance map~(Eqn.~\ref{eqn:cov}) of the electron kinetic spectrum following 790~eV ionization of N$_2$O. (C) Partial covariance map~(Eqn.~\ref{eqn:pcov}) demonstrating suppression of spurious correlations in panel (B). This reveals correlation between photoelectrons from ionization of the nitrogen $K$-shell (${\sim}$380-400eV) and Auger-Meitner emission (${\sim}$340-360eV), highlighted with the blue rectangle. 
        The inset shows the average spectral profile of the incoming x-ray pulse, which produces the bimodal structure of the nitrogen $K$-shell photoelectron spectrum.}
        \label{fig:EEcov}
    \end{figure}

    To demonstrate an advantage of a high collection efficiency electron spectrometer such as our MBES, we performed an electron/electron covariance measurement on an N$_2$O target ionized by $790$~eV x-rays.
    Ionization of the nitrogen $K$-shell produced photoelectrons with a kinetic energy of ${\sim}390$~eV.
    The core-ionized molecule decayed \textit{via} Auger-Meitner emission, leading to the emission of a second electron with a kinetic energy of ${\sim}350$~eV. %
    A schematic of this decay process is shown in Fig.~\ref{fig:EEcov}(A).
    We measure the kinetic energy spectrum of the resultant electrons with our MBES, using a retardation voltage of $250$~V applied to the electrostatic lens.
    For this measurement, we made use of the electrons detected on the outer anode to improve the effective collection efficiency of the MBES.
    We calculated the autocovariance of the electron kinetic energy spectrum across ${\sim}40,000$ XFEL shots~\cite{frasinski_covariance_2016}:
    \begin{equation}
        \mbox{Cov}[\vec{Y},\vec{X}]=\left\langle \vec{Y}\cdot\vec{X} \right\rangle - \left\langle \vec{Y} \right\rangle\left\langle \vec{X} \right\rangle,
        \label{eqn:cov}
    \end{equation}
    where $\vec{X}$ is the electron kinetic energy spectrum, $\vec{Y}=\vec{X}^{\mbox{T}}$, and operator $\langle\cdot\rangle$ indicates an average across the measured laser shots. 
    Fig.~\ref{fig:EEcov}~(B) shows a small region of the resultant covariance map.
    This map is dominated by strong covariance islands between electrons of kinetic energy ${\sim}330-400$ and ${\sim}265$~eV, which lie directly below the horizontal black line at $280$ eV.
    Electrons with kinetic energy ${\sim}$380--400 eV correspond to direct ionization from the N$1s$ shell, while electrons with energy ${\sim}$340--360~eV are produced by nitrogen Auger-Meitner emission.
    The electrons at ${\sim}265$~eV are produced by ionization from the oxygen $K$-shell.
    
    In XFEL measurements, the shot-to-shot fluctuations of the spectral profile and pulse energy of the incoming x-ray pulse can produce spurious covariance islands which do not reflect the correlated emission of two photoproducts from the same ionization pathway.
    To remove these uninteresting correlations, here we calculate the multivariate partial covariance:
    \begin{equation}
        \mbox{pCov}[\vec{Y}, \vec{X}; \vec{I}] = \mbox{Cov}[\vec{Y}, \vec{X}] - \mbox{Cov}[\vec{Y}, \vec{I}]\left(\mbox{Cov}[\vec{I}^{T}, \vec{I}]\right)^{-1}\mbox{Cov}[\vec{I}, \vec{X}],
        \label{eqn:pcov}
    \end{equation}
    where $\vec{I}$ is the single-shot x-ray spectrum and $\left(\mbox{Cov}[\vec{I}^{T}, \vec{I}]\right)^{-1}$ is the inverse of the autocorrelation of the spectrum. 
    This partial covariance map is shown in Fig.~\ref{fig:EEcov}~(B).
    In this map we observe that the spurious correlation between the electrons at ${\sim}330-400$ and ${\sim}265$~eV has been suppressed.
    We also observe a strong positive correlation away from the diagonal of the map, indicating that the emission of electrons with ${\sim}390$~eV of kinetic energy is correlated with the emission of the $360$~eV Auger-Meitner electrons as illustrated in panel (A).
    This is a direct experimental observation of the photoelectron/Auger-Meitner electron emission process following $K$-shell ionization of N$_2$O.
    The kinetic energy structure of the Auger-Meitner emission results from the population of different final states in the N$_2$O dication.
    We note that the nitrogen photoelectron features have a dual peak structure.
    This is the result of the average spectral intensity of the incoming x-ray spectrum, which has a double-peak structure as shown in an inset in each panel.

    \subsection{Spatial Sensitivity of MBES} \label{spatial_sensitivity}

 \begin{figure*}
            \centering
            \includegraphics[width =\textwidth,height =\textheight,keepaspectratio]{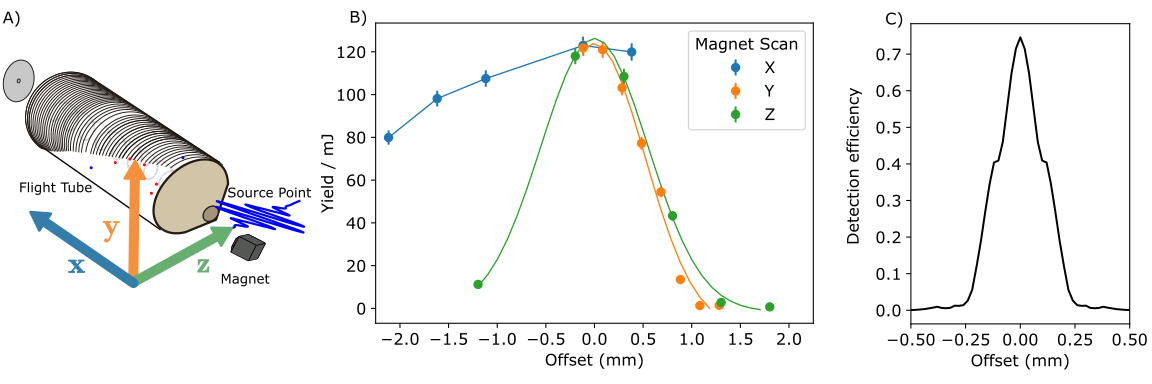}
            \caption{(A) Schematic illustrating the coordinate definitions. \textit{x}: time-of-flight axis of the magnetic bottle and x-ray polarization direction, \textit{y}: axis of the diffusive gas jet; \textit{z}: direction of x-ray propagation.(B) Electron yield from N$_2$O ionized by $516$~eV x-rays as a function of magnet alignment according to coordinate definitions in panel (A). (C) Simulated collection efficiency of the MBES for a point source of electrons displaced from the interaction point by the $yz$ plane.}
            \label{fig:yield_magnet}
        \end{figure*}
        
    In contrast to open area spectrometers~(such as traditional VMI~\cite{eppink_velocity_1997} or COLTRIMS~\cite{dorner_cold_2000}) time-of-flight spectrometers collect electrons from a localized region. 
    In a magnetic bottle spectrometer the volume of the collection region is small, due to the strong spatial localization of the magnetic field.
    This spatial selectivity offers the opportunity to perform photoionization measurements which are highly selective to different positions along the focus of the ionizing radiation.
    
    The collection efficiency of the MBES is sensitive to the alignment between the permanent magnet, the flight tube and the electron source.
    We can optimize the spectrometer alignment using the three-axis manipulator on which the magnet is mounted.
    We plot the dependence of the electron yield on the magnet position in Fig.~\ref{fig:yield_magnet}~(B), which is sensitive to the spatial selectivity of the MBES.
    This measurement was performed using electrons with ${\sim}100$~eV kinetic energy produced by nitrogen $K$-shell ionization of N$_2$O by ${\sim}510$~eV x-rays.
    No retardation potential was applied to the lens.
    We observe a nearly equal dependence on collection efficiency for magnet motion along the axes perpendicular to the flight tube ($y$ and $z$-axes), with a collection length of ${\sim}1$~mm.     
    The range of motion we are able to scan is limited by spatial constraints on the position of the magnet.
    The reduced sensitivity along the axis of the spectrometer~($x$-axis) is expected. 
    Moving the magnet in this direction can affect the collection efficiency of electrons depending on their initial direction of emission \cite{cheshnovsky_magnetic_1987}.     
    The measurement shown in Fig.~\ref{fig:yield_magnet} is sensitive to the spatial selectivity of the MBES, but does not allow for its direct characterization.
    This is because we were unable to change the alignment between the flight tube and the electron source.
    Moreover, at the beam parameters available for the experiment the electron signal was produced by linear x-ray interactions.
    As a result, our effective electron source was highly extended along the direction of beam propagation.
    Nonetheless, we are able to investigate the spatial selectivity of the MBES in simulation.
    For similar parameters as the measurement shown in Fig. \ref{fig:yield_magnet}(B) (103~eV KE electrons, 0~V retardation), we simulate the spatial selectivity of the MBES by displacing a point source of electrons, emitted in a uniform distribution across the unit sphere, in a MBES where the magnet and flight tube are fully aligned.
    The direction of the source point displacement is perpendicular to the flight tube axis, which corresponds to any direction in the $yz$ plane illustrated in panel (A).
    The simulated dependence of detection efficiency on the source point position is shown in panel (C).
    We observe that in a well-aligned MBES, it is possible to strongly limit the collected electrons to a source with a spatial extent around 100~$\mu$m.
     
    \subsection{Energy Resolution} \label{energy_resolution}
        
\begin{figure}
        \centering
        \includegraphics[width =0.5\textwidth,height =\textheight,keepaspectratio]{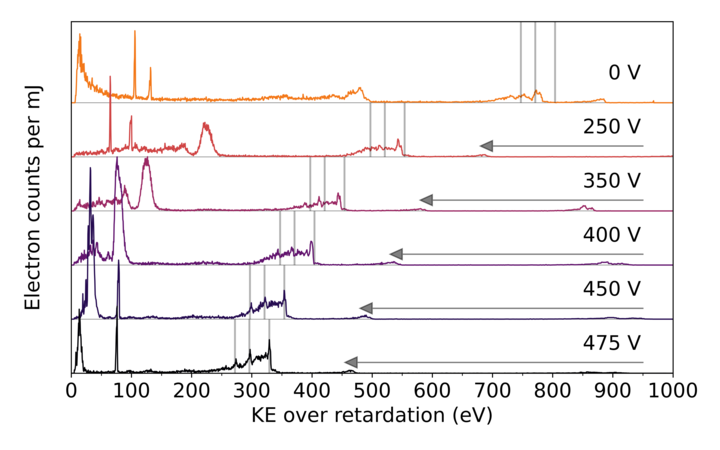}
        \caption{
        Photoelectron kinetic energy spectrum of neon recorded at a photon energy of 1.35~keV. 
        The three primary peaks of the Ne $KLL$ Auger–Meitner emission spectrum become increasingly better resolved as the voltage applied to the electrostatic retardation lens is increased.
        The three dashed lines correspond to previously measured values of these features~\cite{krause_multiple_1970}.}
        \label{neon_retardation}
    \end{figure}

    To investigate the energy resolution of the MBES and characterize the effect of the retardation lens, we performed a systematic scan of the retardation voltage.
    A higher retardation voltage is expected to produce higher energy resolution, because the energy resolution of an MBES $\nicefrac{\delta E}{E}$ is approximately linearly proportional to $\nicefrac{\Delta T}{T}$ \cite{kruit_magnetic_1983}.

    Fig. \ref{neon_retardation} shows the measured photoemission kinetic energy spectrum of neon gas ionized by ${\sim}1.35$~keV x-rays.
    We record this spectrum at six different retardation voltages, as indicated on the figure.
    The neon $KLL$ Auger-Meitner emission has three dominant features separated by ${\sim}$30 eV, corresponding to different final states of the neon dication \cite{krause_multiple_1970}.
    The expected energetic position of these three features is indicated by the vertical dotted lines.
    We observe that increasing the retardation voltage improves the energy resolution of the spectrometer: at higher retardation voltages the three primary $KLL$ emission lines appear more clearly separated.
    
    To quantify the resolving power of the MBES, we measured the widths of the photoionization features produced by $K$-shell ionization of N$_2$O as a function of retardation voltage.
    This measurement was performed using x-rays with a central photon energy of ${\sim}516$~eV.
    The target molecule was chosen because the nitrogen $K$-shell photoemission spectrum of N$_2$O is well-studied and consists of two features corresponding to ionization of the central and terminal nitrogen sites at $412.5$~eV and $408.5$~eV, respectively \cite{griffiths_doubly_1991}.
    
    The SASE FEL has a broad (${\sim}$5~eV) spectral bandwidth and inherent shot-to-shot fluctuations in the spectral profile, which can significantly degrade spectroscopic resolution.
    To mitigate the effect of the SASE bandwidth and spectral jitter, and isolate the resolution of the MBES, we employed spectral domain ghost imaging~\cite{li_attosecond_2022,li_two-dimensional_2021}, as described in the appendix.

    \begin{figure*}
        \centering
        \includegraphics[width =\textwidth,height =\textheight,keepaspectratio]{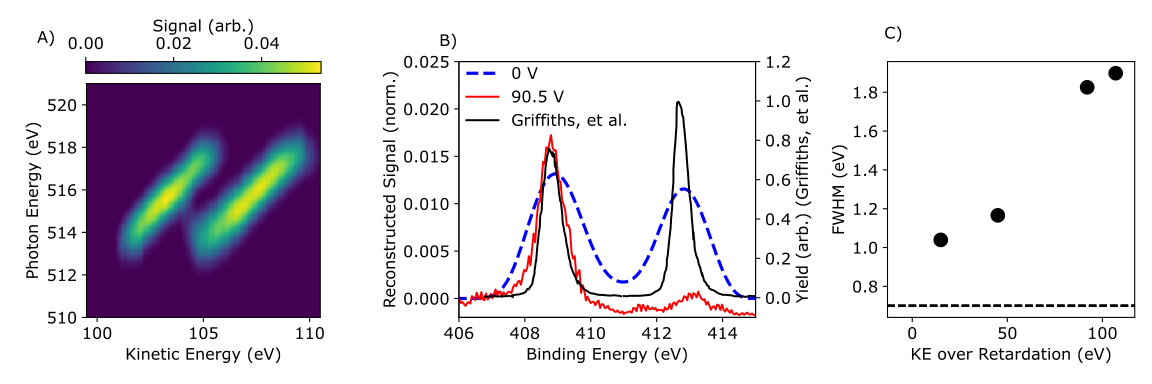}
        \caption{ 
        (A) Dispersion plot of photoelectron kinetic energy $vs.$ photon energy for the N(1s) photoionization of N$_{2}$O, obtained using spectral domain ghost imaging.
        (B) Binding energy spectrum for N(1s) photoionization of N$_{2}$O obtained from dispersion plot in panel A.
        The measured peak width decreases with increasing retardation voltage. Black line shows previous high-resolution measurements made with a synchrotron source from \cite{griffiths_doubly_1991}. 
        The change in relative intensities at larger retardation voltages is a result of the MBES transmission function, see Fig. \ref{fig:det_eff_lineout}.
        (C) Widths corresponding to gaussian fits of the terminal photoline centered at 408.5~eV binding energy. The dashed black line at 0.70~eV corresponds to the measured width from reference \cite{griffiths_doubly_1991}.} 
        \label{spook_recon}
    \end{figure*}

    An example of the two-dimensional map of photoelectron kinetic energy \textit{vs.} incoming photon energy is presented in Fig.~\ref{spook_recon}~(A). 
    We extract the kinetic energy spectrum from the two-dimensional map by averaging across all photon energies after accounting for the energy dispersion of the photoelectron.
    The width of the measured photoelectron feature for different retardation values is shown in Fig.\ref{spook_recon} (B) and can be seen to decrease with increasing retardation.
    The MBES detection efficiency at very low values of kinetic energy over retardation is highly suppressed, as shown in Fig.~\ref{fig:det_eff_lineout}.
    This effect results in a significant decrease in yield for the lower energy photoelectrons at a retardation of 90.5~V.
    We quantify the resolution by fitting a gaussian curve to the photoline of the terminal nitrogen site at 107 eV.
    The dependence of this width on retardation is shown in Fig. \ref{spook_recon} (C).
    The results show that when there is no retardation voltage applied, it is still possible to discern the 4 eV separation between the central and terminal lines thanks to the long flight tube.
    Nonetheless by increasing the retardation voltage, the kinetic energy resolution can be significantly increased.

\section{Energy-Dependent Modulation in Electron Transmission Efficiency}
    \begin{figure*}
        \centering
        \includegraphics[width=0.7\textwidth,keepaspectratio]{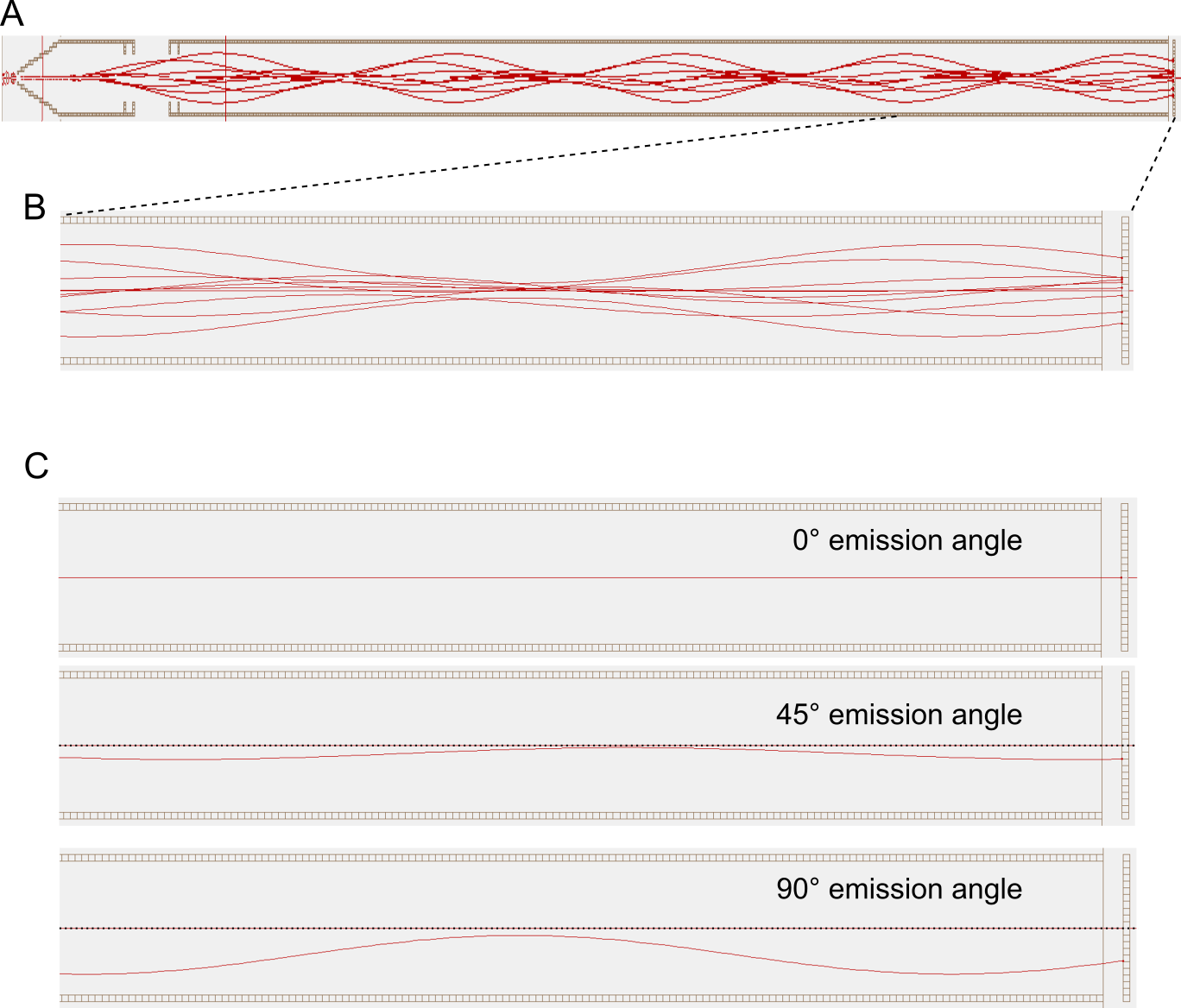}
        \caption{ (A) SIMION simulation of trajectories of ${\sim}$300~eV electrons with different initial emission directions through the MBES. The retardation voltage is 0~V and the solenoid current has been reduced from the standard operation parameters to amplify the radius of the cyclotron oscillation. (B) Zoom-in of the trajectories at the end of the flight tube, close to the electron detector. (C) Dependence of electron trajectories through the flight tube on the angle of direction of initial emission, relative to the flight tube axis. The dotted  black line shows the central axis of flight tube.}
        \label{fig:sausaging-simion}
    \end{figure*}
Finally, we discuss an artifact that is common to magnetic bottle electron spectrometers, namely a strongly energy-dependent modulation of the electron transmission function which is often amplified when a retarding field is applied to the electrostatic lens.
This artifact is a result of the betatron oscillations of the electrons as they travel through the magnetic-field of the flight tube and propagate toward the detector~\cite{barba_magnetic_2020}.
The cyclotron radius of these oscillations, 
\begin{equation}
    r_{C}=\frac{\sqrt{2m_eKE}}{e_0|\vec{B}|}\sin{\theta_{f}},
    \label{eqn:cyc_rad}
\end{equation}
depends on the electron kinetic energy $KE$, the strength of the magnetic field $\vec{B}$, and the angle of the electron velocity relative to the flight tube axis, $\theta_{f}$.
The effect of this cyclotron motion is to induce a periodic modulation in the transverse distribution of electron trajectories as they move along the flight tube.
The spatial period of the modulation, i.e. the distance between points where the electron spread is maximal, depends on the kinetic energy of the electron~\cite{barba_magnetic_2020}:  
\begin{equation}
    l_o(KE)=2\pi\frac{\sqrt{2m_eKE}}{e_0|\vec{B}|}\cos{\theta_f}.
    \label{eqn:saus_len}
\end{equation}
The size of the electron spatial distribution at the detector is therefore a function of the velocity of the electrons. 
If the flight tube is an integer multiple of $l_o$, then the spatial extent of the electrons is quite small at the detector plane.
In the other limiting case, if the flight tube length is a half-integer multiple of $l_o$ and the spatial extend of the electron distribution is maximal at the detector plane.
The energy-dependent oscillations have consequences for the operation of a MBES: electrons arriving at the detector plane with radial displacement larger than the detector can miss the detector and fail to be detected.
Thus, the electron detection efficiency of the MBES is periodically modulated with electron time-of-flight.
In our spectrometer, there may be an additional decrease in transmission efficiency due to electrons impinging on the plates of the electrostatic retardation lens.
This additional effect should be particularly pronounced at high retardation voltages, due to fringe fields from the electrostatic lens that act perpendicular to the time-of-flight axis.

This oscillatory behavior of the electron trajectories is illustrated in the simulations shown in Fig.~\ref{fig:sausaging-simion}.
We plot ten trajectories of electrons emitted in arbitrary directions uniformly sampled on a sphere, with ${\sim}$300~eV initial kinetic energy.
The dependence of the cyclotron radius on the direction of the electron velocity described in Eq. \ref{eqn:cyc_rad} manifests as the dependence of the trajectory on the initial emission angle, as shown in panel (C).
This effect is often colloquially referred to as `sausaging', because the three-dimensional shape of the electron trajectories through the flight tube is reminiscent of the shape of a link of sausages.

\begin{figure*}
        \centering
        \includegraphics[width =\textwidth,height =\textheight,keepaspectratio]{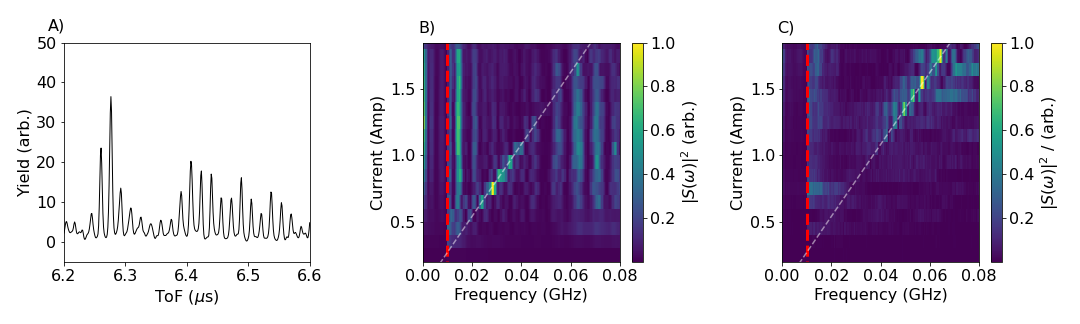}
        \caption{
        (A) Time-of-flight (ToF) spectrum for photoelectrons produced by ionization of N$_2$O by $520$~eV x-rays, measured on the inner anode at a solenoid current of 2~A, showing a strong periodic modulation in detection efficiency.
        (B) Fourier transform of the ToF spectra as a function of applied solenoid current for the outer anode.
        (C) Fourier transform of the ToF spectra as a function of applied solenoid current for the inner anode.
        For frequency values greater than 0.01 GHz, the color scales in panels (B) and (C) have been multiplied by 50 and 25, respectively.}
        \label{sausaging_ft}
    \end{figure*}

This oscillatory collection efficiency is observed in experiment in the measured electron time-of-flight spectra of x-ray~($520$~eV) ionization of N$_2$O molecules, shown in Fig.~\ref{sausaging_ft}~(A).
The strong periodic modulation of electron detection efficiency with the electron time-of-flight is apparent in the spectrum recorded on the inner anode of the detector. 
According to Eqn.~\ref{eqn:saus_len}, the modulation frequency should increase with increasing magnetic field, and thus increasing solenoid current. 
This effect can be seen in Fig. \ref{sausaging_ft}~(B)~and~(C), where the Fourier transform of the electron time-of-flight spectra recorded on the outer ((B)) and inner ((C)) sections of the detector is shown as a function of the current applied to the solenoid coil.
There is a dominant Fourier component, indicated by the white dashed-line, whose frequency increases linearly with solenoid current in accordance with the well-known dependence of cyclotron frequency on magnetic field strength.
Below ${\sim}1.1$~A, the modulation is particularly pronounced on the outer section of the detector~(outer anode), while above this value the effect dominates the inner anode.
This is a result of the geometry of the two detectors: as we increase the solenoid current, the electron trajectories flatten to the central axis of the flight tube and the modulation amplitude grows stronger on the inner anode and less prominent on the outer.

\begin{figure*}
         \centering
         \includegraphics[width =\textwidth,height =\textheight,keepaspectratio]{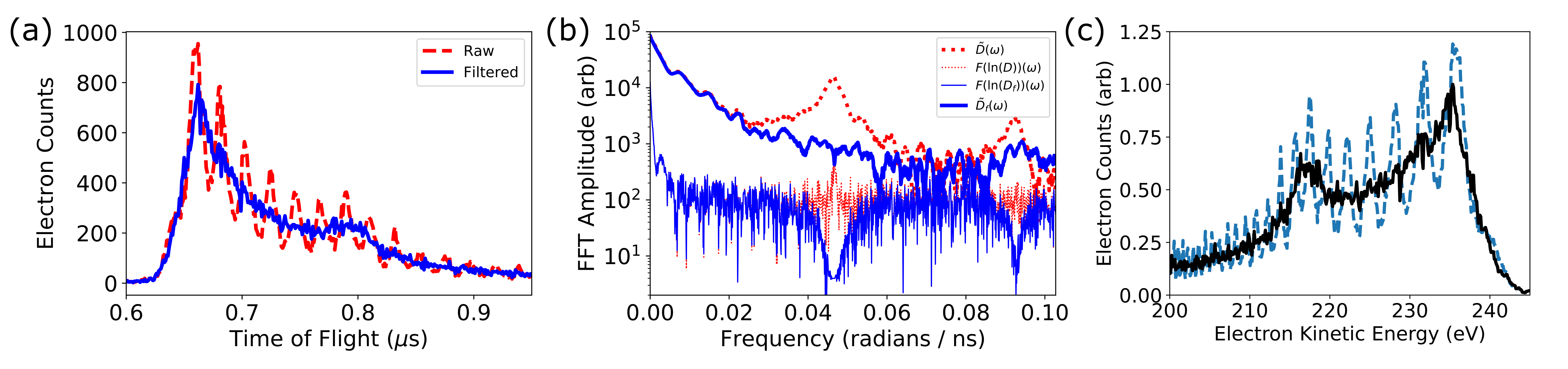}
         \caption{Demonstration of the filtering algorithm to remove structure caused by betatron motion of electrons in the magnetic bottle for the x-ray valence ionization spectrum of para-aminophenol.
         (A) Time-of-flight representation with the raw spectrum showing the artefact before (dashed red) and after (blue) applying the filtering algorithm. 
         (B) FFT power spectrum at different points in the filtering algorithm. Red lines show the spectrum of the raw data before (tick dotted red line) and after (thin dotted red line) taking the logarithm, i.e. $D(\omega)$ and $\mathcal{F}(\ln(D))(\omega)$. Blue lines show the spectrum of the log-filtered data before (thick blue line) and after (thin blue line) exponentiating, i.e. $F(\ln{D_{f}(\omega)})$ and $\tilde{D}_{f}(\omega)$.
         (C) Data in the electron kinetic energy representation before (blue dashed line) and after (black line) the filtering algorithm.}
         \label{fig:desausage_algorithm}
     \end{figure*}

One solution to the periodic modulation in electron detection efficiency is to increase the solenoid current to reduce the cyclotron radius such that all electrons fall on the detector.
However, increasing the solenoid current to the required value may be technically challenging, and also results in a decrease of electron energy resolution~\cite{kruit_magnetic_1983}.
Instead we developed an analytical method to remove this artifact in data post-processing~\cite{oneal_tracking_2022}.
We demonstrate this routine using the electron spectrum recorded by our MBES following the ionization of gas-phase para-aminophenol at ${\sim}$252 eV.
The time-of-flight spectrum was recorded with a retardation of $190$~V and is shown in Fig.~\ref{fig:desausage_algorithm}~(A) before~(dashed red) and after~(dashed blue) removal of this artifact.

To remove the artifact, we treat the measured electron spectrum ~$D(t)$ as a product of the true electron spectrum~$S(t)$ and the transmission function that encodes the periodic features, which we approximate as~$T_s(t) = 1-a \cos(\omega_s t)$, where $\omega_s$ is the period of oscillation in the transmission function. 
The measured data can then be written as
\begin{equation}
    D(t) = S(t)T_s(t) = S(t)(1 - a \cos(\omega_s t)).
\end{equation}
Filtering out the modulation frequency~($\omega_{s}$) from this transmission function with a Fourier filter will result in data loss because the Fourier transform of~$D(t)$ is given by the convolution rather than the sum,
\begin{equation}\label{eq:fftd}
    \tilde{D}(\omega) = (\tilde{S} \ast \tilde{T}_s)(\omega) \propto 2\tilde{S}(\omega) - a \left(\tilde{S}(\omega - \omega_s) + \tilde{S}(\omega + \omega_s)\right).
\end{equation}
Attempts to filter out the shifted copies of~$\tilde{S}$ are likely to modify the extracted spectrum. 

To avoid this problem, we first take the logarithm of the spectrum, which separates the ground truth spectrum and transmission function into a sum, which can be more easily separated by Fourier decomposition:
\begin{equation}\label{eq:lnd}
    \ln(D(t)) = \ln(S(t)T_s(t)) = \ln(S(t)) + \ln(1-a \cos(\omega t)).
\end{equation}
The final term can be written as a Fourier series as,    
\begin{align}\nonumber
    \mathcal{F}(\ln(D))(\omega) = & \mathcal{F}(\ln(S))(\omega) - \ln \left( 1 + \frac{a^2}{4} \right) \delta(\omega) \\ \nonumber
    & + a (\delta(\omega - \omega_s) + \delta(\omega + \omega_s)) \\
    & + \frac{a^2}{4} (\delta(\omega - 2 \omega_s) + \delta(\omega + 2 \omega_s)) + \mathcal{O}(a^3).
\end{align}

The broadening at the harmonic ~$ 0.045 \nicefrac{rad}{ns}$ due to the convolution in Eq.~\ref{eq:fftd} is replaced by a narrowed peak after applying the logarithm, as shown in Fig.~\ref{fig:desausage_algorithm}~(B).
To recover the signal~$S(t)$, we filter out the harmonics of $\omega_s$ from the log-scaled spectra using Butterworth filters to generate the frequency-domain filtered data $\mathcal{F}(\ln(D))(\omega)$. 
To recover the filtered data in real space $D_{f}(t)$, we apply an inverse Fourier transform and then exponentiate.
This method of examining the Fourier transform of the logarithm has been used to study situations where the Fourier transform contains periodic peaks, because it maps the convolution of two functions onto their sums~\cite{childers_cepstrum_1977}.

\begin{figure*}
        \centering
        \includegraphics[width =0.8\textwidth,height =\textheight,keepaspectratio]{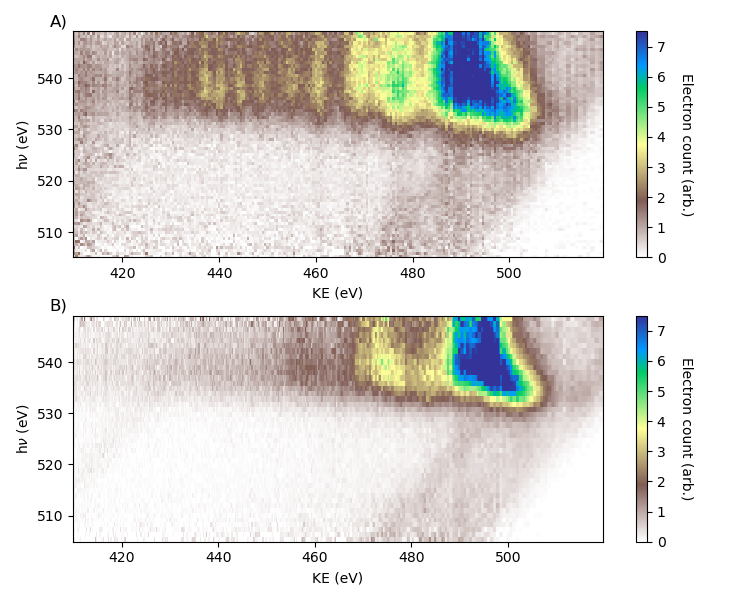}
        \caption{Resonant oxygen Auger-Meitner spectra of para-aminophenol before (A) and after (B) the filtering algorithm to remove the periodic modulation in detection efficiency. The $x$-axes shows the electron kinetic energy spectrum measured with a retardation of ${\sim}$400~eV and the $y$-axes show the x-ray central photon energy. Between x-ray photon energies of 530--535~eV we observe resonant Auger-Meitner emission following resonant oxygen 1s $
        \rightarrow$ valence excitation. At photon energies above 535~eV, the spectrum converges to regular oxygen $KLL$ Auger-Meitner emission. We also observe dispersive lines due to x-ray valence ionization.}
        \label{desausage}
    \end{figure*}

We also implemented the log-space Fourier filter algorithm on a two-dimensional measurement of the resonant oxygen Auger-Meitner emission from para-aminophenol. 
The X-ray photon energy was scanned over the oxygen $K$-edge region from ${\sim}$505 to ${\sim}$550~eV, and the high-energy electron emission spectrum was recorded with $400$~V applied to the electrostatic lens. 
The resonant Auger-Meitner map (incoming x-ray photon energy \textit{vs.} outgoing photoelectron energy) is shown in Fig.~\ref{desausage}.
The photoelectrons produced by x-ray ionization of the valence shell show a clear linear dispersion, and the resonant Auger-Meitner emission after promotion of an electron from the oxygen $1s$ to the $2p\pi^\ast$ orbital is clear at ${\sim}$532~eV.
Above the oxygen $1s$ ionization potential, the electron emission converges to the normal $KLL$ Auger-Meitner emission.

In the left panel, the periodic modulation is clearly seen and is most apparent in the resonant and normal Auger-Meitner emission at higher photon energies.
We note that this modulation does not disperse with photon energy because it is a function of electron time-of-flight only.
Removal of this periodic modulation has been performed by an enhanced application of the algorithm described above. 
To eliminate the possibility of distorting the two-dimensional map, a single transmission function $T(t)$ was developed and then applied uniformly to each photon energy bin as
\begin{equation}
    D_{f}(t,h\nu)=\nicefrac{D(t,h\nu)}{T(t)}
\end{equation}
where $D_{f}(t,h\nu)$ denotes the two-dimensional filtered data.
To compute the transmission function, the raw data is summed over photon energy and the 1D curve is filtered. 
By dividing the unfiltered and filtered data, a robust transmission function can be developed as
\begin{equation}
    T(t) = \nicefrac{D_{all}(t)}{D_{all,f}(t)}
\end{equation}

This transmission function is more robust compared to treating each photon energy bin separately, because the filter is nonlinear.
By increasing the statistics when calculating the transmission function, we can more effectively filter the periodic transmission function.

\section{Conclusion}
In this paper, we have presented the design and performance of a MBES for x-ray photoelectron spectroscopy.
Our spectrometer is installed at the TMO endstation of the Linac Coherent Light Source and is available to users of the facility.
It is equipped with an einzel-stack retardation lenses and a segmented anode, which enables high kinetic-energy resolution~($\delta E/E \sim 1 \%$) measurements of high energy (several hundred eV) photoelectrons.
The segmented anode also enables the simultaneous collection of data in two different modes: high collection efficiency, using electrons detected on both anodes, and high resolution, using only electrons detected on the inner anode.
We achieve high energy x-ray photoelectron spectroscopy measurements when coupled to a noisy x-ray free-electron laser by employing spectral domain ghost imaging, and we achieve covariance measurements of correlated photoelectron/Auger-Meitner electron emission from N$_2$O.
We present a robust analysis procedure to correct for the well-known periodic modulation in detection efficiency due to cyclotron motion of the electron in the MBES flight tube.
Upgrades to the instrument will include the integration of an ion time-of-flight spectrometer for concurrent ion/electron spectroscopy.
Our spectrometer can currently be used for a variety of user experiments, and will be capable of operating in the MHz regime with the the upgraded high-repetition-rate LCLS-II source.

\section*{Acknowledgements}
We are grateful to Jan Metje, Markus G{\"u}hr and Richard Squibb for useful discussions on the design of the MBES and to John Pennachio and Denise Welch for excellent technical support.
KDB and DR acknowledges the support of the Department of Energy, Chemical Sciences, Geosciences and Biosciences Division, Office of Basic Energy Sciences, Office of Science, grant No. DE-FG02-86ER13491.
KB acknowledges the support of the US Department of Energy, Office of Science, Office of Workforce Development for Teachers and Scientists, Office of Science Graduate Student Research (SCGSR) program. The SCGSR program is administered by the Oak Ridge Institute for Science and Education (ORISE) for the DOE. ORISE is managed by ORAU under contract number DE-SC0014664
DR acknowledges the hospitality and support of SLAC during his sabbatical.
NB acknowledges the support of the Department of Energy, Chemical Sciences, Geosciences and Biosciences Division, Office of Basic Energy Sciences, Office of Science, grant No. DE-SC0012376

\section*{Author Declarations}

\section*{Conflict of Interest}

The authors have no conflicts to disclose.

\subsection*{Data Availability}
The data that support the findings of this study are available from the corresponding author upon reasonable request.

% \addbibresource{references.bib}
% \addbibresource{.bib}

\section{Appendix}

\subsection{Spectral Domain Ghost Imaging}
    Spectral domain ghost imaging exploits the correlation between electron and x-ray photon spectra on a single-shot basis to extract the spectral response of the sample under investigation \cite{driver_attosecond_2020,li_two-dimensional_2021,li_time-resolved_2021,wang_photon_2023}.
    In the single photon regime, the correlation of the photoelectron spectra with the photon flux of the incident x-ray pulse can be represented as a system of linear equations $Ax=b$. 
    $A$ and $b$ are matrices whose columns index the spectral bins of the photon and electron energy spectra respectively, and whose rows correspond to different shots.
    The unknown partial ionization cross section is encoded in the $x$ matrix.
    Due to the inherent measurement noise in $A$ and $b$, determining $x$ $via$ least-square optimization is typically a numerically unstable operation.
    Therefore, we solve for $x$ by minimizing the following cost function, which includes regularization parameters which enforce known qualities of the solution $x$:
    
\begin{equation}
    ||Ax-b||_{2}^{2}+\lambda_{1}||x||_{1}+\lambda_{2}||Lx||_{2}^{2}+Ind_{+}(x) 
    \label{eqn:ax-b}
\end{equation}
    
    Here, the $\lambda$ are hyperparameters that weigh the different regularization terms.
    For the measurements presented in this work, the first term imposes sparsity defined by the $L_1$-norm $\lambda_{1}$ and the second term imposes smoothness $\lambda_{2}$, defined by the laplacian operator.
    The last term imposes non-negativity in $x$.
    We employed a quadratic programming optimization \cite{wang_fdgi_URL,osqp} to minimize equation 13
    and obtain the optimal value of $x$.
    We choose the values of the hyperparameters by selecting appropriate values along the L-hypersurface \cite{belge_simultaneous_1998,belge_efficient_2002}.
    For all measurements, the spectral response $x$ was robust to the values of the hyperparameters over several orders of magnitude.

\bibliography{references.bib, references_additional.bib}

\end{document}

%% file: authors.tex
%\author*[1,2]{\fnm{First} \sur{Author}}\email{iauthor@gmail.com}

\author*[1,2]{\fnm{Kurtis} \sur{Borne}} \email{kborne@phys.ksu.edu}
\author[3,4]{\fnm{Jordan~T.} \sur{O'Neal}}
\author[3,5]{\fnm{Jun} \sur{Wang}}
\author[2,3,5]{\fnm{Erik} \sur{Isele}}
\author[2]{\fnm{Razib} \sur{Obaid}}
\author[6]{\fnm{Nora} \sur{Berrah}}
\author[2]{\fnm{Xinxin} \sur{Cheng}}
\author[3,4,5]{\fnm{Philip~H.} \sur{Bucksbaum}}
\author[2]{\fnm{Justin} \sur{James}}
\author[2,3]{\fnm{Andrei} \sur{Kamalov}}
\author[2,3]{\fnm{Kirk~A.} \sur{Larsen}}
\author[2]{\fnm{Xiang} \sur{Li}}
\author[2]{\fnm{Ming-Fu} \sur{Lin}}
\author[2,3]{\fnm{Yusong} \sur{Liu}}
\author[3]{\fnm{Agostino} \sur{Marinelli}}
\author[2]{\fnm{Adam} \sur{Summers}}
\author[3,4]{\fnm{Emily} \sur{Thierstein}}
\author[2,3]{\fnm{Thomas} \sur{Wolf}}
\author[1,2,3]{\fnm{Daniel} \sur{Rolles}}
\author[2]{\fnm{Peter} \sur{Walter}}
\author[2,3]{\fnm{James~P.} \sur{Cryan}}
\author*[2,3]{\fnm{Taran} \sur{Driver}} \email{tdriver@stanford.edu}

%\affil*[1]{\orgdiv{Department}, \orgname{Organization}, \orgaddress{\street{Street}, \city{City}, \postcode{100190}, \state{State}, \country{Country}}}
\affil[1]{\orgdiv{J.R. Macdonald Laboratory, Department of Physics}, 
    \orgname{Kansas State University}, 
    \orgaddress{\city{Manhattan} \postcode{66506}, \state{Kansas}, \country{USA}}}
\affil[2]{ 
    \orgname{SLAC National Accelerator Laboratory}, 
    \orgaddress{ \city{Menlo Park} \postcode{94025}, \state{California}, \country{USA}}}
\affil[3]{\orgdiv{Stanford Pulse Institute}, 
    \orgname{SLAC National Accelerator Laboratory}, 
    \orgaddress{ \city{Menlo Park} \postcode{94025}, \state{California}, \country{USA}}}

\affil[4]{\orgdiv{Department of Physics}, 
    \orgname{Stanford University}, 
    \orgaddress{, \city{Stanford}, \postcode{94305}, \state{California}, \country{USA}}}

\affil[5]{\orgdiv{Department of Applied Physics}, 
    \orgname{Stanford University}, 
    \orgaddress{, \city{Stanford}, \postcode{94305}, \state{California}, \country{USA}}}

\affil[6]{\orgdiv{Department of Physics}, 
    \orgname{University of Connecticut}, 
    \orgaddress{ \city{Storrs}, \postcode{}, \state{Connecticut}, \country{USA}}}